# A study of polarization compensation for quantum networks

**Matej Peranić[1*], Marcus Clark[2,5], Rui Wang[3], Sima Bahrani[3], Obada Alia[3], Sören Wengerowsky[4], Anton Radman[1], Martin Lončarić[1], Mario Stipčević[1], John Rarity[2], Reza Nejabati[3], Siddarth Koduru Joshi[2]**

[1] *Photonics and Quantum Optics Research Unit, Center of Excellence for Advanced Materials and Sensing Devices, Ruđer Bošković Institute, Zagreb, Croatia*

[2] *Quantum Engineering Technology Labs, H. H. Wills Physics Laboratory & Department of Electrical and Electronic Engineering, University of Bristol, Bristol, United Kingdom*

[3] *High Performance Networks Group, School of Computer Science, Electrical & Electronic Engineering and Engineering Maths (SCEEM), University of Bristol, Bristol, United Kingdom*

[4] *ICFO - Institut de Ciencies Fotoniques, The Barcelona Institute of Science and Technology, Castelldefels (Barcelona), Spain*

[5] *Quantum Engineering Centre of Doctoral training, NSQI, University of Bristol, Bristol, United Kingdom*

*Corresponding author: Matej.Peranic@irb.hr

## Abstract

The information-theoretic unconditional security offered by quantum key distribution has spurred the development of larger quantum communication networks. However, as these networks grow so does the strong need to reduce complexity and overheads. Polarization-based entanglement distribution networks are a promising approach due to their scalability and no need for trusted nodes. Nevertheless, they are only viable if the birefringence of all-optical distribution fibres in the network is compensated to preserve the polarization-based quantum state. The brute force approach would require a few hundred fibre polarization controllers for even a moderately sized network. Instead, we propose and investigate four different realizations of polarization compensation schemes that can be used in quantum networks. We compare them based on the type of reference signals, complexity, effort, level of disruption to network operations and performance on a four-user quantum network.

**Introduction**

Quantum key distribution (QKD) provides provable security for protocols used to exchange encryption keys and hence encrypted messages between multiple users [1-3]. Typically, most implementations have focused on individual QKD links consisting of two users. To establish individual quantum communication links, several research groups have used a wide variety of qubit encodings [4-6]. One common method, especially for in-fibre QKD, is polarization encoding [7-9]. Using polarization states of photons in QKD simplifies the end-user hardware in comparison to time-bin encoding since this encoding method does not require interferometers. However, any polarization encoding scheme suffers from the birefringence of the optical fibres used and will not work without some form of polarization compensation to ensure that the polarization axes from the source are faithfully transmitted to the receiver [10, 11]. In this way, the polarization basis system is identical at the source and the receiver. Typically, active polarization compensation uses specialized hardware that sends its own bright classical signal through the optical fibre [12-14]. However, the bright classical light signals used will cause insurmountable noise on the single photon level quantum signals. Thus, there is a need for adequate polarization compensation schemes that work well with polarization-based QKD protocols [1, 2]. These schemes have to avoid the use of bright classical light and reduce the power level of the reference signal to single photons. We note that procedures that may be easy to implement in simple 2-user quantum links may not necessarily be scalable as we start building large-scale and heavily interconnected quantum networks. In that sense, methods that work for 2-user links may improve individual links in the networks, but not the network as a whole. This can lead to a high number of controllers needed for networks with more than a couple of users. For example, the power level of bright classical light can be reduced to a single-photon level using an optical attenuator, but it requires a classical signal to follow the same path as a quantum signal. This leads to the compensation of links in the network one by one for each pair of users which requires $2k$ controllers for $k$ links. On the other hand, we show that auxiliary classical light as a reference signal is not needed and that minimization of Quantum Bit Error Rate (QBER) can be used in the compensation process in quantum networks. Not only that this approach reduces the number of controllers needed to $k$ for $k$ links but also it does not require downtime of the network.

We present polarization compensation schemes with different types of reference signals on a four-user quantum network and compare their benefits and drawbacks in the context of QKD. We assess their performances in the entanglement-based quantum communication network with a polarization-entangled photon pair source and wavelength division multiplexing technique.

We evaluated polarization compensation methods based on the ease of implementation, resources needed, time taken for compensation of each fibre, the amount of user participation needed and whether causing disruption to the quantum network service. After describing the experimental setup, we present each of the four realizations used and how they were implemented. In the discussion, we critically discuss the performance of each method against the above criteria.

**Experimental setup**

Lately, significant efforts have been made in the implementation of QKD in networks in which each user is connected with all other users at the same time [15-17]. One common way to create such a full-mesh network is wavelength multiplexing which allows expanding the bandwidth capacity of the network without the need for additional fibres [9]. To connect four users in a full-mesh quantum network, we have used a type-0 polarization-entangled photon pair source with a spectrum of ~ 60 nm full width at half maximum (FWHM) in a Sagnac interferometer configuration [18, 19]. The pump laser at 775.06 nm produces signal and idler photons with wavelengths symmetrically distributed around the central wavelength of 1550.12 nm in the process of spontaneous parametric down-conversion. This process [20] occurs in the 5 cm long Magnesium Oxide doped periodically poled Lithium Niobate (MgO:ppLN) bulk crystal with a polling period of 19.2 µm. Combining clockwise and counter clockwise contributions in a Sagnac interferometer at the polarization beamsplitter, we create a maximally entangled state $|\phi^+\rangle = \frac{1}{\sqrt{2}}(|H\rangle_1|H\rangle_2 + |V\rangle_1|V\rangle_2)$ for each pair of photons whose wavelengths are symmetrical about the central wavelength [15]. The central wavelength corresponds to the ITU channel 34 according to the ITU-T G.694.1 recommendation, and the selected pairs of signal and idler photons can be distributed to different users using a quantum reconfigurable add-drop multiplexer (q-ROADM) [21, 22]. Q-ROADM consists of a wavelength de-multiplexer (DEMUX) that divides the broadband entangled photon spectrum into 30 × 100GHz ITU channels, fibre polarization controllers (FPC), an optical fibre switch (OFS) and a multiplexer (MUX) that combines previously divided entangled photons in a single fibre such that every user receives three wavelengths to form a fully connected network (Fig. 1, b). Using q-ROADM, we introduce active switching to provide flexibility and dynamic control over network configuration. Each user in a network is provided with a polarization analysis module (PAM) that enables polarization analysis in two mutually unbiased orthonormal bases. The two outputs of each PAM are connected to superconducting nanowire detectors from Photon Spot with detection efficiencies between 70% and 90% and jitter between 80 ps and 60 ps that are further connected to a time-tagger unit (Swabian Time-Tagger Ultra). Although the detectors are polarization dependent, it does not affect the process since the approach with minimization of the number of photons only requires that minimum is achieved, no matter what the exact number of photons in the minimum is, as described in the next section. The distances between users in the network are 1.6 km, and the measured loss varies from 8.1 dB to 13 dB, depending on the link. These values include loss from fibre transmissions, q-ROADM, PAMs and detectors.

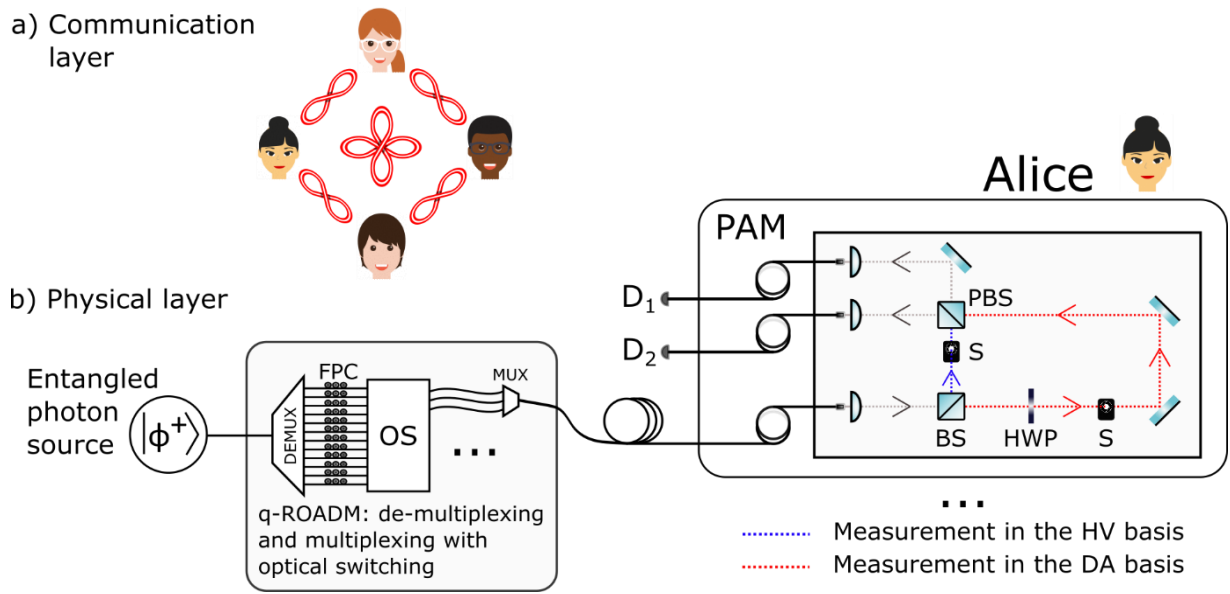

Figure 1. a) Communication layer of the four-user full-mesh network. Every pair of users share a bipartite entangled state as represented by each individual infinity symbol. b) Physical layer of the network consists of a polarization-entangled photon source, q-ROADM (consisting of fibre polarization controllers (FPC), optical switch (OS), de-multiplexer DEMUX and multiplexer MUX) and polarization analysis module (PAM). Each user has a PAM consisting of a beamsplitter (BS), polarization beamsplitter (PBS), half waveplate (HWP), shutters (S), and mirrors. Single-photon detectors are depicted as $D_1$ and $D_2$. Solid lines depict optical fibres and dashed lines free-space path of photons.

**Canonical method using an auxiliary laser for polarization compensation**

- Realization with manual polarization controllers

In a physical layer, users in the network are connected with optical fibres (Fig. 1, b). Unavoidable mechanical stress due to external conditions, present along any realistic deployed optical fibre, will result in a transformation of an incident polarization and increase the Quantum Bit Error Rate (QBER). This undesired transformation can be compensated for using adjustable polarization controllers. The canonical method implies sending an auxiliary laser light with selectable and predefined polarization states (usually switching between two different mutually unbiased bases) through the same optical fibre that will later carry the quantum signal to the users. In a wavelength multiplexed quantum network, this reference signal must be of the same wavelength as the intended quantum signal. Hence, we use a tunable laser as our source. Further, to use the same single-photon detectors for the polarization compensation and for QKD we use a variable optical attenuator (VOA) to reduce the laser power to a suitable level. This saves the time and effort that would otherwise be needed to switch from single-photon detectors to photodiodes for signal measurement at the output (Fig. 2).

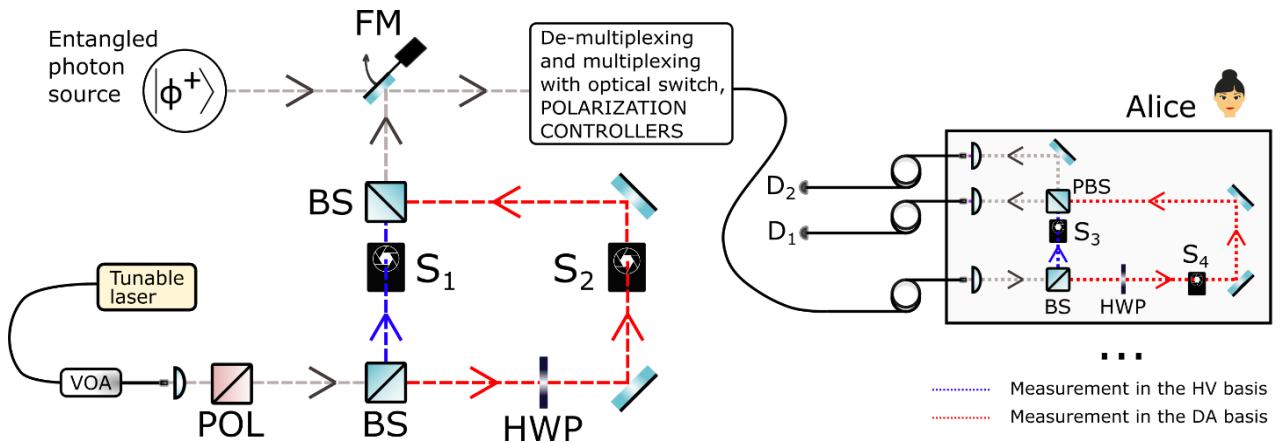

Figure 2. Setup for polarization compensation using predefined photon polarization states. The laser power is controlled with the variable optical attenuator (VOA). The polarization of the photons entering the setup is defined using the polarizer (POL), in our case Wollaston prism. The flip mirror (FM) is used to switch between the signal from the setup for polarization compensation and the signal from the source of polarization-entangled photon pairs. Mechanical shutters are depicted as $S_1$, $S_2$, $S_3$ and $S_4$, beamsplitters as BS, polarization beamsplitter as PBS, half waveplates as HWP and single-photon detectors as $D_1$ and $D_2$ Solid lines are describing optical fibres and dashed lines free-space path of photons.

Laser light from a tunable laser is prepared in the H state (horizontal linear polarization) using a Wollaston prism (POL in fig. 2). Photons transmitted through the first BS in the setup for predefining the polarization state end up in the D state (diagonal linear polarization) due to rotation on a HWP and photons reflected on the first BS remain in the H state. We choose to send photons of the either H or D polarization state of the classical light to the users by closing one of the corresponding shutters ($S_1$ and $S_2$ in fig 2.). They experience the same disturbance as the quantum signal and end up entering PAMs in a random state before the implementation of polarisation compensation. Besides polarization, we must take care of the wavelengths too. Since each pair of users is connected with different pairs of ITU channels, in the process of compensation we also must send photons of corresponding wavelengths. The procedure for compensation requires sending photons of one of two polarization states (H or D in our setup) and measuring the polarization state received at the PAM. Then, the measurement is performed at the detector corresponding to the orthogonal state in the same basis since it is generally preferable to align to a minimum rather than a maximum. This means that we send H and measure in V or send D and measure in A. Optical shutters in the transmitter and receivers are used to ensure that the measurement basis always corresponds to the correct setting based on the well-defined sent polarization state. We note that the shutters are only needed for the polarization compensation steps and are both left open (closed) on the receivers (transmitter) during the QKD protocol. Compensation in one basis is done when the minimum value attainable is observed with the corresponding detector. After compensation in one basis, we send the other polarization state and compensate in this basis as well. With iterative alternations between both transmitted polarization states, we are able to find a common position of the fibre polarization compensation paddles (Fig. 3, left) that results in the lowest values for the V detector (when we send H) and the A detector (when we send D). The results of our measurements based on the sample of 206 compensations on a full-mesh four-user network show that it takes 14 min average per link to achieve average polarization visibility of (98.17 ± 0.04)% after compensation. Our previous research [23] showed polarization stability that corresponds to the average standard deviation of QBER of 0.6% after compensation with manual controllers over a period of 10.8 days. This period includes cooling cycles of superconducting detectors every 24 hours over which the network was not running but it fully recovered its functionality after cycling. Although the network cannot operate at the same time when compensation is conducted, this scheme provides high polarization visibilities and the possibility for fine adjustments in both bases.

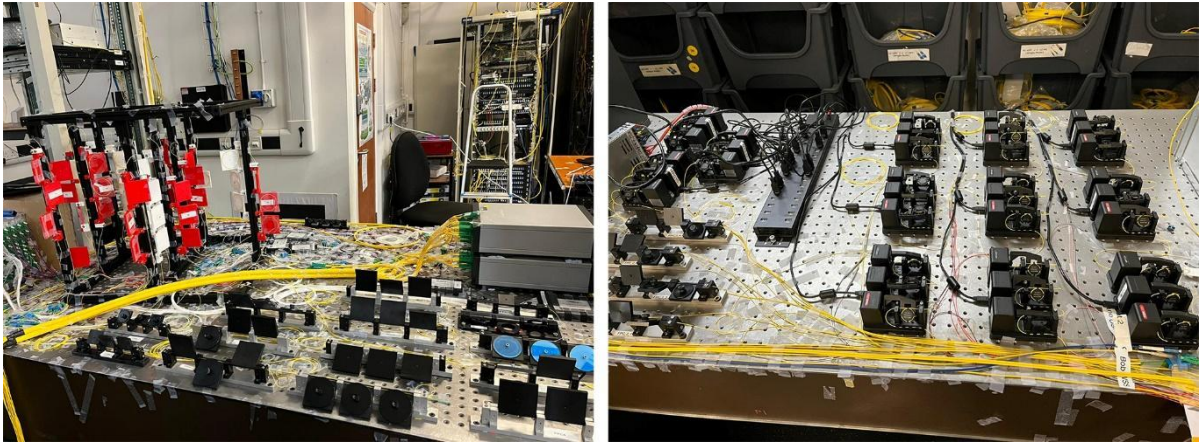

Figure 3. Left: Photography of a part of the experimental setup with manual polarization controllers, Right: Photography of a part of the experimental setup with motorized polarization controllers

- Realization with motorized polarization controllers

Although widely used, manual polarization controllers suffer from limitations induced by human factors. On the other hand, motorized polarization controllers (MPC) offer reproducibility and are easy to use [12, 24]. Here, we have implemented MPCs (Fig. 3, right) to our network with the algorithm that maximizes polarization visibility above a certain threshold. Besides threshold polarization visibility values in each base, the user can define the global polarization visibility threshold that he would like to achieve, the initial angle, and the step size that depends on the value of visibility. It is natural to take larger steps when being far off the optimum value and to refine steps closer to the polarization visibility threshold. As described in Fig. 4, in the first step, our algorithm finds the paddle with the highest impact on the polarization visibility value, positions it to maximize polarization visibility, and excludes that paddle. In the next step, the algorithm checks if polarization visibility is higher than the predetermined threshold value in that basis and, if that condition is not satisfied, finds the second paddle with the highest impact. Further rotations of those two paddles are enabling us to get the result above the threshold value. This algorithm is very natural in the sense that it follows steps that one would take to compensate polarization using manual polarization controllers. To compare, using motorized polarization controllers with our algorithm results in similar polarization visibility (above 98%) as manual compensation, but faster (8 min). Considering that other methods have shown to be even faster, for further investigations we recommend combining the best of both worlds, reproducibility and automation of the algorithm using MPCs with the blinking scheme or with the possibility of avoiding the disruption of the network with the QBER minimization. In our experiment, all MPCs start from the same position and move for 10° as their initial angle and try to achieve 95% polarization visibility for the HV base, 98% polarization visibility for the DA base, and 95% global polarization visibility as an average between HV base and DA base. Even though the polarization visibility in one basis might be lower than the threshold value in that basis, if the global polarization visibility is larger than the global threshold value, the algorithm stops. The algorithm will run up to four times in each base before it reduces the threshold value that it is trying to achieve for 0.2%. Also, it will try to switch to another base up to 10 times before it stops if the global threshold is not achieved. The consequence of high threshold polarization visibility values is the possibility that those values won't be achieved in the first try, but on the other hand, we have achieved the lowest contribution to QBER after polarization compensation (Table 1.) in comparison to manual fibre polarization controllers and "blinking scheme".

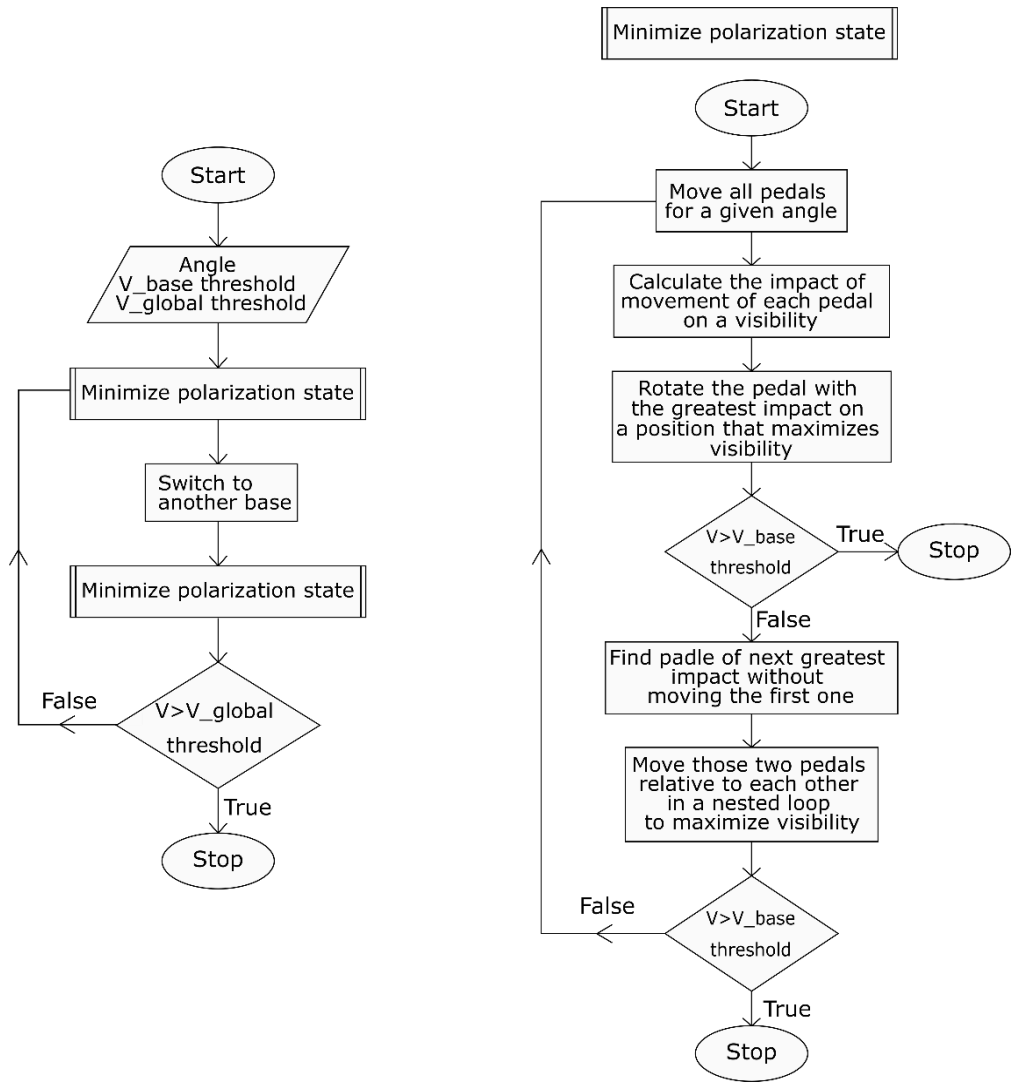

Figure 4. Flowchart of the algorithm used to control MPCs

- Realization with simultaneous polarization compensation in both polarization bases – "blinking scheme"

A realization of a canonical method using manual polarization controllers requires iterative steps of changing the predefined state of a reference signal before compensation. This process can be time-consuming since the iterative procedure does not necessarily converge. However, if one could send and receive polarization states from both bases (i.e. H state from HV basis, and D state from DA basis, alternately) for a short time, it could be possible to compensate in both bases "simultaneously". In both arms of the setup for polarization state preparation, we have shutters that can be open, closed, or blinking with some frequency. The same pair of shutters can be found on the users' PAM modules (Fig. 2). In our experiment, both pairs of shutters were working in a "blinking" mode with an integration time of 0.3 s per base, enabling us to track the polarization visibility of both bases simultaneously. We notice that there is a compromise between how fast shutters open and close, and the experimental practicality. Faster blinking of shutters would give a better average but it leads to the mixing of different basis due to imperfect shutters synchronization. Experimentally, we noticed this effect by achieving lower maximum polarization visibility (in both bases) while working in blinking mode with lower integration times than 0.3 s compared to the manual realization where shutters are opening and closing arms one by one. Although this scheme also requires downtime of the network, it is much faster than manual realization. Compensation done on 24 links shows an average polarization visibility value of (97.6 ± 0.2)% in 6 minutes per link. In this way, we have reduced the network's downtime by more than half with a similar network performance compared with the previously described manual realization.

**Minimization of QBER**

For entanglement-based QKD protocols [2, 25], QBER below 11% is required to ensure a positive secret key rate [26]. As previously demonstrated for two users [27-30], it is possible to use a quantum signal in the process of polarization compensation. Here, we propose the first implementation of the scheme with the minimization of QBER for quantum networks. Unlike the previously described methods where adding an $n$-th user in the network would require compensation of variations for $2(n – 1)$ new links, with this method each new user needs to compensate only fibres that are connecting him/her to the source. In this way, communication links with other users will be polarization compensated leaving the rest of the network intact (Fig. 5). Although this method requires finding the exact delay between users to calculate the QBER correctly, its big advantage is that the process of compensation can be done while the network is active and without any additional hardware. This could play a crucial role in real-life implementations. After finding the delay between a pair of users, QBER is calculated from a temporal cross-correlation histogram [15, 25]. While monitoring its live value, QBER was minimized using manual controllers. However, we note that it is impossible to have continuous monitoring of compensation since the signal can not be used for compensation and key generation simultaneously.

We have done compensation on a running four-user network with live QBER monitoring on 13 links and results show entanglement fidelity of (93.2 ± 0.8)% for what takes around 2 min on average per link.

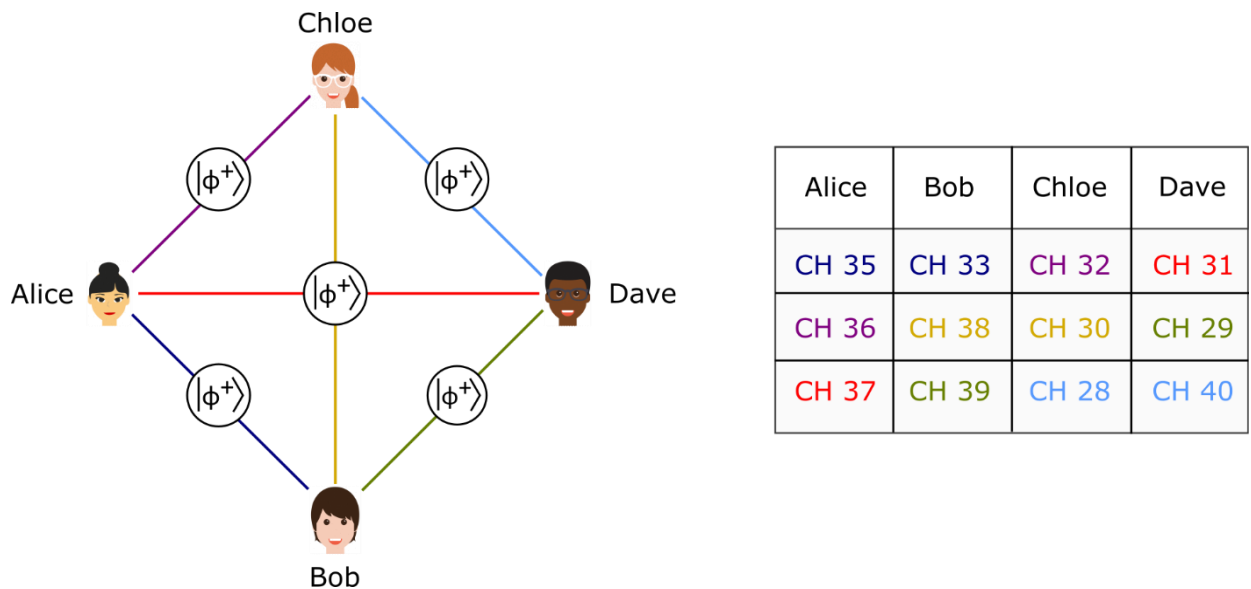

Figure 5. QBER minimization scheme effectively corresponds to having a source that produces entangled photon pairs at wavelengths corresponding to ITU channels shown in the table between each pair of users. The colours in the table represent different photon pairs that are distributed among the users, while channel numbers correspond to wavelengths symmetrically distributed around the central wavelength of 1550.12 nm (corresponding to channel 34).

**Discussion**

We have successfully applied all presented schemes for polarization compensation in a quantum communication network, but we note that some of them have advantages that could be crucial when scaling quantum networks to a high number of users. Besides the method with QBER minimization, other methods require $2k$ FPCs for $k$ links. Since every $n$-th new user in a full-mesh network based on multiplexing needs to establish $2(n-1)$ new links, it is important to use an appropriate method. Even more importantly, since minimization of QBER does not require downtime, network management can adjust polarization visibility to keep it above a certain threshold even during the communication process. In addition, it significantly reduces the time needed for the process of compensation for which it stands out among the other methods. In Table 1. we present a summary of evaluated methods and their realizations with the results of measurements. We note that while the QBER-based method is clearly preferred, it requires a high-fidelity state from the entangled photon pair source and a high coincidence rate. If the fidelity of the state needs to be tuned, then it becomes necessary to have at least two users sharing one link whose fibres have been polarization compensated by one of the other methods. If the coincidence rate is low, a longer time is needed to obtain a useful QBER value. Further, the QBER-based polarization compensation scheme is not suitable as a diagnostic tool, since it does not differentiate between errors caused by the birefringence in the optical fibre and the fidelity of the source of entangled photons. We have shown that there is no real performance difference between manual fibre polarization compensation and motorised ones when using the paddle-based controllers (i.e. rotating loops of fibre). Our simple deterministic algorithm presented here is sufficient to rapidly compensate the fibres. The main limitation to their operating speed was their rotation speed and the readout time of the detector counts. Using mechanical shutters in the "blinking scheme" is clearly advantageous because it eliminates the extra manual steps of changing the emitted and measured states. However, it is important that the shutters blink in tandem with each other. Asynchronous operation of the shutters was found to enable a fraction of the light to be sent or measured in the wrong polarization state/basis. Given that other methods have proven effective, the use of shutters is questionable. However, their use as a diagnostic tool for remotely deployed user modules remains efficacious.

Table 1. Summary of polarization compensation methods for quantum networks

| Method | Canonical with an auxiliary laser | | | Minimization of QBER |
|---|---|---|---|---|
| Realization | Manual fibre polarization controllers | Motorized polarization controllers | „Blinking scheme" using mechanical shutters | Compensation done on one fibre connecting pair of users |
| Figure | 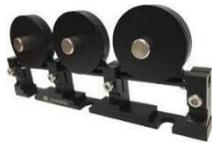 | 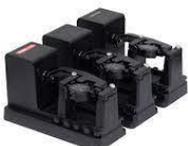 | 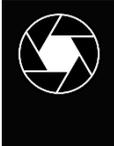 | 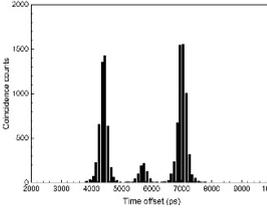 |
| FPCs needed for *k* links | 2*k* | 2*k* | 2*k* | *k* |
| Do not disrupt the whole network | ❌ | ❌ | ❌ | ✅ |
| Calibration signal not needed | ❌ | ❌ | ❌ | ✅ |
| Active changes of basis not needed | ❌ | ❌ | ❌ | ✅ |
| *t* (per link) | ~14 min | ~8 min | ~6 min | ~2 min |
| Polarization visibility after compensation | (98.17 ± 0.04)% | (98.4 ± 0.2)% | (97.6 ± 0.2)% | not applicable |
| Entanglement fidelity | 93.3% (estimated*) | 93.5% (estimated*) | 92.7% (estimated*) | (93.2 ± 0.8)% (measured) |
| Calculated contribution to QBER due to polarization compensation | (0.91 ± 0.02)% | (0.77 ± 0.03)% | (1.18 ± 0.08)% | 0.05% (estimated*) |
| Measured QBER | ranging from 2.7% to 4.0%† | not measured | | (3.4 ± 0.4)% |

*Estimated including 3.35% (average of measured QBERs during calibration measurements) net QBER contribution (i.e. 93.3% entanglement fidelity contribution) from entangled photon pair source, user modules, timing jitter, detectors and polarization compensation
†The actual experimentally measured value depends on the properties of the network link and thus there is a nearly uniform spread of QBER values between the extremes quoted

**Conclusion**

We have demonstrated four realizations of polarization compensation methods that can be used in quantum networks. The promising results of their practical implementation have confirmed that they can be a powerful tool for future large-scale QKD networks. The process of polarization compensation with minimization of QBER can be done while the network is active and utilises only photons from the entangled photon source, which is an important comparative advantage. Implementation of motorized controllers with the QBER method could further reduce the time required for the polarization compensation process and automate it. We did not find an efficient automation solution for the QBER-based method other than a simple grid search algorithm and consequently did not implement it. Future work could investigate Machine Learning algorithms to help compensate polarization changes in the optical fibres based on the QBER value. Also, different technical realizations of the presented methods could be investigated. Importantly, our experiment shows that each logical link within the entanglement distribution layer of the quantum network can be polarization compensated independently and we do not need to compensate each wavelength over each physical link in the network.

**List of references**

1. Bennett CH, Brassard G. International Conference on Computer System and Signal Processing IEEE 1984;175–9.

2. Ekert A. Quantum cryptography based on Bell's theorem. Phys. Rev. Lett. 1991;67:661. doi:10.1103/PhysRevLett.67.661

3. Bennett CH, Brassard G, Mermin ND. Quantum Cryptography without Bell's Theorem. Phys. Rev. Lett. 1992;68:557. doi: 10.1103/PhysRevLett.68.557

4. Boaron A, Korzh B, Houlmann R, Boso G, Rusca G, Gray S et al. Simple 2.5 GHz time-bin quantum key distribution. Appl. Phys. Lett. 2018;112:171108. doi:10.1063/1.5027030

5. Khan IA, Howell JC. Experimental demonstration of high two-photon time-energy entanglement. Phys. Rev. A. 2006;73:031801. doi:10.1103/PhysRevA.73.031801

6. Grosshans F, Van Assche G, Wenger J, B 269 rouri R, Cerf NJ, Grangier P. Quantum key distribution using gaussian-modulated coherent states. Nature. 421, 2003;238-41. doi:10.1038/nature01289

7. Jennewein T, Simon C, Weihs G, Weinfurter H, Zeilinger A. Quantum Cryptography with Entangled Photons. Phys. Rev. Lett. 2000;84:4792. doi:10.1103/PhysRevLett.84.4729

8. Yin J, Li YH, Liao SK, Yang M, Cao Y, Zhang L et al. Entanglement-based secure quantum cryptography over 1,120 kilometers. Nature. 2020;582:501-5. doi:10.1038/s41586-020-2401-y

9. Wengerowsky S, Joshi SK, Steinlechner F. An entanglement-based wavelength-multiplexed quantum communication network. Nature. 2018;564225-8. doi:10.1038/s41586-018-0766-y

10. Noe R, Heidrich H, Hoffmann D. Endless polarization control systems for coherent optics. Journal of Lightwave Technology. 1988;6(7);1199-1208. doi: 10.1109/50.4117

11. Chen J, Wu G, Li Y, Wu E, Zeng H. Active polarization stabilization in optical fibers suitable for quantum key distribution. Opt. Express. 2007;15;17928-17936. doi: 10.1364/OE.15.017928

12. Xavier GB, De Faria GV, Temporão GP, Von der Weid JP. Full polarization control for fiber optical quantum communication systems using polarization encoding. Opt. Express. 2008;16:1867-73. doi:10.1364/OE.16.001867

13. Chen J, Wu G, Li Y, Wu E, Zeng H. Active polarization stabilization in optical fibers suitable for quantum key distribution. Opt. Express. 2007;15(26):17928-36. doi:10.1364/OE.15.017928

14. Chen J, Wu G, Xu L, Gu X, Wu E, Zeng H. Stable quantum key distribution with active polarization control based on time-division multiplexing. New J. Phys. 2009;11:065004. doi:10.1088/1367-2630/11/6/065004

15. Joshi SK, Aktas D, Wengerowsky S, Lončarić M, Neumann SP, Liu B et al. A trusted node-free eight user metropolitan quantum communication network. Science advances. 2020;6(36):8. doi:10.1126/sciadv.aba0959

16. Liu X, Xue R, Huang Y, Zhang W. Fully Connected Entanglement-based Quantum Communication Network without Trusted Node. Optical Fiber Communication Conference (OFC). 2021

17. Qi Z, Li Y, Huang Y, Feng J, Zheng Y, Chen X. A 15-user quantum secure direct communication network. Light Sci Appl. 2021;10:183. doi:10.1038/s41377-021-00634-2

18. Kim T, Fiorentino M, Wong FNC. Phase-stable source of polarization-entangled photons using a polarization Sagnac interferometer. Phys. Rev. A. 2006;73(1); 012316. doi:10.1103/PhysRevA.73.012316

19. Lim HC, Yoshizawa A, Tsuchida H, Kikuchi K. Broadband source of telecom-band polarization-entangled photon-pairs for wavelength-multiplexed entanglement distribution. Opt. Express. 2008;16;16052-16057. doi: 10.1364/OE.16.016052

20. Schneeloch J, Knarr SH, Bogorin DF, Levangie ML, Tison CC, Frank R et al. Introduction to the absolute brightness and number statistics in spontaneous parametric down-conversion. Journal of Optics. 2019;21(4);043501. doi: 10.1088/2040-8986/ab05a8

21. Wang R, Joshi SK, Kanellos GT, Aktas D, Rarity J, Nejabati R, Simeonidou D. AI-Enabled Large-Scale Entanglement Distribution Quantum Networks. Optical Fiber Communication Conference (OFC) 2021, Dong P, Kani J, Xie C, Casellas R, Cole C, Li M, eds., OSA Technical Digest (Optica Publishing Group, 2021), paper Tu1I.4. doi: 10.1364/OFC.2021.Tu1I.4

22. Wang R, Alia O, Clark MJ, Bahrani S, Joshi SK, Aktas D et al. A Dynamic Multi-Protocol Entanglement Distribution Quantum Network. Optical Fiber Communication Conference (OFC) 2022, Matsuo S, Plant D, Wey JS, Fludger C, Ryf R, Simeonidou D, eds., Technical Digest Series (Optica Publishing Group, 2022), paper Th3D.3. doi: 10.1364/OFC.2022.Th3D.3

23. Clark MJ, Alia O, Wang R, Bahrani S, Peranić M, Aktas D et al. Entanglement distribution quantum networking within deployed telecommunications fibre-optic infrastructure. Proc. SPIE 12335, Quantum Technology: Driving Commercialisation of an Enabling Science III, 123350E (2023). doi:10.1117/12.2645095

24. Agnesi C, Avesani M, Stanco A, Villoresi P, Vallone G. All-fiber autocompensating polarization encoder for quantum key distribution. Optics Express. 2019;44(10):2398-401. doi:10.1364/OL.44.002398

25. Xiongfeng M, Fred F C-H, Hoi-Kwong L. Quantum key distribution with entangled photon sources, Phys. Rev. A, 2007;76(1);012307, doi: 10.1103/PhysRevA.76.012307

26. Gobby C, Yuan ZL, Shields AJ. Quantum key distribution over 122 km of standard telecom fiber. Appl. Phys. Lett. 2004;84:3762. doi:10.1063/1.1738173

27. Shi Y, Poh HS, Ling A, Kurtseifer C. Fibre polarization state compensation in entanglement-based quantum key distribution. Opt. Express. 2021;29:37075-80. doi: 10.1364/OE.437896

28. Ding YY, Chen W, Chen H, Wang C, Li Y-P, Wang S et al. Polarization-basis tracking scheme for quantum key distribution using revealed sifted key bits. Opt. Lett. 2017;42:1023-6. doi:10.1364/OL.42.001023

29. Neumann SP, Buchner A, Bulla L, Bohmann M, Ursin R. Continuous entanglement distribution over a transnational 248 km fiber link. Nat Commun. 2022;13;6134. doi:10.1038/s41467-022-33919-0

30. Ramos MF, Silva NA, Muga NJ, Pinto AN. Full polarization random drift compensation method for quantum communication. Opt. Express. 2022;30:6907-20. doi:10.1364/OE445228

**Abbreviations**

**QKD**: Quantum Key Distribution

**QBER:** Quantum Bit Error Rate

**FWHM:** Full Width at Half Maximum

**MgO:ppLN:** Magnesium Oxide doped periodically poled Lithium Niobate

**Q-ROADM**: Quantum Reconfigurable Add-Drop Multiplexer

**DEMUX**: De-Multiplexer

**FPC**: Fibre Polarization Controller

**OS**: Optical Switch

**MUX:** Multiplexer

**PAM**: Polarization Analysis Module

**BS**: Beamsplitter

**PBS:** Polarization Beamsplitter

**HWP:** Half-Wave Plate

**VOA:** Variable Optical Attenuator

**POL**: Polarizer

**WP:** Wollaston Prism

**FM:** Flip Mirror

**MPC:** Motorized Polarization Controller

**Declarations**

**Availability of data and materials**

The datasets used and analysed in this study are available from the corresponding author upon request.

**Competing interests**

The authors declare that they have no competing interests.

**Funding**

The research leading to this work has received funding from United Kingdom Research and Innovation's (UKRI) Engineering and Physical Science Research Council (EPSRC) Quantum Communications Hub (Grant Nos. EP/M013472/1, EP/T001011/1), British Scholarship Trust, Agency for Mobility and EU Programmes, Croatian Science Foundation, HRZZ grant No. IPS-2020-1-2616 and Croatian Ministry of Science and Education, MSE grant No. KK.01.1.1.01.0001.

**Authors' contributions**

The source of entangled photon pairs was built by SKJ and optimized by MC and MP. The polarization analysis modules were built by MP under the supervision of ML. The idea for the research was discussed between MP, ML, SKJ and SW. The data was collected, analysed, and interpreted by MP, MC, RW, SB, and OA under the supervision of ML, MS and SKJ. The software and the electronics were developed by AR, SKJ, MC, and RW. The manuscript was written by MP and reviewed by all authors. All authors discussed the results and commented on the manuscript. The financing of the study was ensured by MS, JR, RN and SKJ.

**Acknowledgements**

We thank Djeylan Aktas and Sebastian Philipp Neumann for helpful discussions. We are grateful to colleagues from the Workshop of Physical Chemistry Division at Ruđer Bošković Institute in Zagreb for great contribution in building PAMs.